# The Management and Integration of Biomedical Knowledge: Application in the Health-e-Child Project (Position Paper)

E. Jimenez-Ruiz[1], R. Berlanga[1], I. Sanz[1], R. McClatchey[2],
R. Danger[1], D. Manset[3], J. Paraire[3], A. Rios[3]

[1]University Jaume I, Castellon, Spain
{ejimenez, berlanga, isanz, danger}@uji.es
[2]CCS Research Centre, University of the West of England (UWE), Bristol, UK
Richard.McClatchey@uwe.ac.uk
[3]Maat Gknowledge, Valencia, Spain
{dmanset, jparaire, arios}@maat-g.com

**Abstract.** The Health-e-Child project aims to develop an integrated healthcare platform for European paediatrics. In order to achieve a comprehensive view of children's health, a complex integration of biomedical data, information, and knowledge is necessary. Ontologies will be used to formally define this domain knowledge and will form the basis for the medical knowledge management system. This paper introduces an innovative methodology for the vertical integration of biomedical knowledge. This approach will be largely clinician-centered and will enable the definition of *ontology fragments*, connections between them (*semantic bridges*) and enriched ontology fragments (*views*). The strategy for the specification and capture of fragments, bridges and views is outlined with preliminary examples demonstrated in the collection of biomedical information from hospital databases, biomedical ontologies, and biomedical public databases.

## 1 Introduction

The Health-e-Child (HeC) project [1] aims for the construction of a Grid-based service-oriented environment to manipulate distributed and shared heterogeneous biomedical data and knowledge sources. This biomedical knowledge repository will allow clinicians to access, analyze, evaluate, enhance and exchange integrated biomedical information and will also enable the use of integrated decision support and knowledge discovery systems. The biomedical information sources will cover six distinct levels (also referred to as vertical levels), classified as molecular, cellular, tissue, organ, individual, population, and will focus on paediatrics, in particular, on some carefully selected representative diseases in three different categories: paediatric heart diseases, inflammatory diseases, and brain tumours.

The HeC project will have several medical institutions contributing diverse biomedical data for the different vertical levels. It is likely that data sources for each level will have different schemata, using different software packages with varying types of access controls. In order to bring these disparate sources together it is necessary to identify the core entities for each level, to build an intermediary data model per level to capture the entities' structures, and to unify these level data models. A set of biomedical ontologies will be used to formally express the HeC medical domain with the mentioned *vertical abstraction levels*. This paper also introduces the concept of an Integrated Disease Knowledge Model (IDKM), which captures the core entities for each vertical level and provides the valid concepts for a particular disease.

## 1.1 Issues in Biomedical Data Integration

Data source integration has been a traditional research issue in the database community. The main goal of an integrated database system is to allow users to access a set of distributed and heterogeneous databases in a homogeneous manner. The key aspect of data integration is the definition of a global schema, but it is worth pointing out that we must distinguish between three kinds of global schemata: the database schemata, the conceptual schemata and domain ontologies. The first describes the data types with which information is locally stored and queried; the second generalizes these schemata by using a more expressive data model like UML (TAMBIS [2] and SEMEDA [3] follow this approach). Finally, domain ontologies describe the concepts and properties involved in a domain (such as Biomedicine) independently of any data model, facilitating the expression of the semantics of the application resources (e.g. via semantic annotation) as well as reasoning about them.

Medical research has a long tradition in unifying terminological concepts and taxonomies (e.g. through the Unified Medical Language System (UMLS) [4]), and in using ontologies to represent and query them in medical information systems. Recently, several approaches to integrating medical and bioinformatics public databases have been ontology based (e.g. ONTOFUSION [5]). However, new issues and challenges arise from the introduction of domain ontologies when integrating information sources. Firstly, many domain ontologies in Biomedicine do not cover completely the requirements of specific applications. Moreover, these concepts can involve different abstraction levels (e.g. molecular, organ, disease, etc.) that can be in the same or in different domain ontologies. Secondly, domain ontologies are normally rather large, resulting in two main effects: users find them hard to use for annotating and querying information sources and only a subset of those are used by system applications. Finally, in current integration approaches, it is necessary to manually map the existing data sources to domain concepts, which implies a bottleneck in large distributed scenarios.

This paper mainly focuses on the two first issues: managing multiple domain ontologies and presenting personalised ontology views to end-users and applications involved in an integrated biomedical information system. The proposed approach consists of a new ontology-based methodology that spans the entire integration process. This methodology relies on both the definition of ontology-based views and their construction from domain ontology fragments.

## 2   Methodology for the Vertical Knowledge Integration

The most important aspect of HeC, in contrast to current biomedical integration projects (e.g. INFOGENMED [7], MyGrid [6], TAMBIS, etc.), is to integrate patient information according to disease models, instead of integrating public biomedical databases. An Integrated Disease Knowledge Model (IDKM) is proposed as a solution to specify the concepts of particular diseases, taking into account all the biomedical abstraction layers. Patient-centric information collected in the hospitals will be semantically annotated in terms of a particular IDKM. Following the Description Logic terminology, the distributed repositories that store the patient semantic annotations are called ABoxes (or Assertional Boxes).

The methodology presented here provides the necessary mechanisms to build IDKMs from well-known biomedical ontologies and public databases. Most simply stated, it enables building ontology-based *views* from consistent fragments of biomedical ontologies, which are interrelated by means of so-called *semantic bridges*. Each ontology fragment is intended to capture the main concepts involved in a disease for a particular abstraction layer (e.g. genetic, organ, etc.). Bridges perform the actual vertical integration, where they relate selected elements of an abstraction layer to those of a more abstract one. In this methodology, bridges can be found explicitly in the biomedical ontologies (e.g. NCI, GO, FMA, etc.) or implicitly in text-rich public biomedical databases (e.g. UNIPROT, OMIM, EMBL, etc.).

Constructing such IDKM models requires going through the following stages (see Figure 1 for a graphical representation of methodology steps):

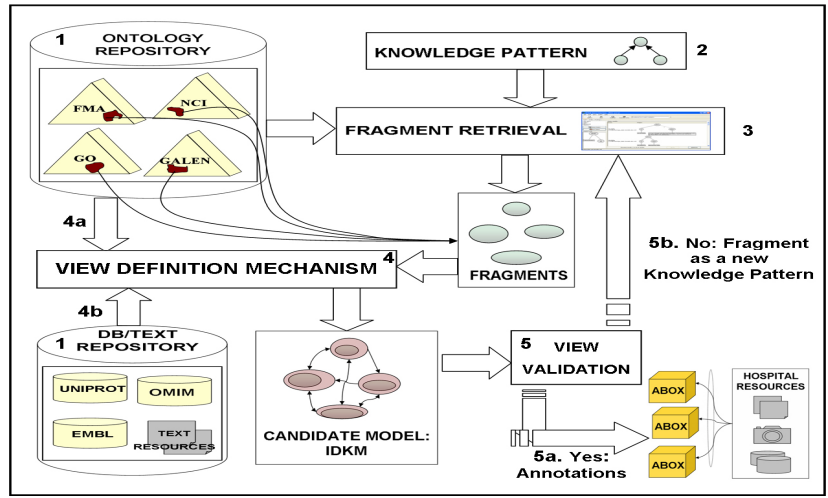

**Fig. 1.** Global Schema of the Methodology

1. **Creation of a Knowledge Repository.** To apply the presented methodology, a set of well-known domain ontologies and public databases have been collected.
2. **Definition of a Knowledge Pattern.** We start the construction of IDKMs from a knowledge pattern: a set of concepts, a hierarchy of concepts (tree), a graph, etc.

3. **Ontology fragments retrieval.** Candidate regions, with respect to a knowledge pattern, are identified in the ontologies through approximate tree matching.
4. **View Definition Mechanism.**
    a. *Definition of complete views.* Previously introduced fragments are then enriched with other concepts and roles from the ontologies by means of a set of inference rules.
    b. *Connecting view fragments.* Views are merged using mapping techniques and inferring connections (*semantic bridges*) from the public databases.
5. **Validation of Views**: The resulting view will be an IDKM candidate model.
    a. *Annotating.* Patient information collected in hospitals has to be annotated according to validated IDKM concepts and roles. The annotation information (or semantic representation) constitutes the A-Boxes.
    b. **Feedback**. If the view is not sufficiently complete, it can be used as a new knowledge pattern and start again the methodology cycle.

## 3   Retrieving Ontology Fragments with ArHex

For the purposes of this study, the tool ArHex [8] has been adopted to retrieve ontology fragments in order to guide the building of the IDKMs. Thus, starting from a collection of ontologies and a knowledge pattern, users can query the knowledge and progressively construct the required IDKM. However, using multiple ontologies raises the problem of semantic heterogeneity, as different concepts can have similar lexical expressions. In the presented approach these problems have been addressed by the introduction of approximate queries [9]. Basically, an approximate query is a *tree pattern* whose nodes specify which concepts and roles have to be found, and arcs that express the different approximate relationships between them (e.g. parent/child, ancestor/descendant, etc.). The retrieval system provides a list of ontology fragments ranked with respect to a similarity measure that compares candidate regions and patterns. We are currently developing a set of base similarity measures suitable for the HeC project, as well as extending the pure tree-oriented ArHeX indexing engine to support directed acyclic graphs, which are required for more powerful ontology querying facilities.

## 4   Definition of Consistent View Fragments

Obtained ontology fragments cannot be directly used to build a consistent IDKM for several reasons. Firstly, some of the selected ontology fragments can conflict and secondly, ontology fragments are sometimes too small and/or incomplete for an IDKM. Therefore, it is necessary to complete these retrieved fragments and to check possible conflicts between their extensions. Fragments provide information about the context of the query concepts, and help in defining views over an ontology, since they bring more information about neighbour concepts and relations. Thus, the view mechanism can be seen as a technique to enrich, with other concepts and relations, the extracted or identified fragments.

At this point in time, the definition of such views has been achieved through the use of a traversal-based view definition language, called OntoPathView [10]. In this language, views over an ontology consist of the union of a set of traversal queries (paths) and a set of inference rules in order to get closed, consistent and complete views [10].

## 5   Representation of Vertical Levels: Modules and Mappings

The identification of the knowledge represented in the ontologies and the coverage over the identified vertical levels is a crucial aspect in the application of this methodology. Figure 2 illustrates the example of a possible coverage of four biomedical ontologies. In the figure ovals represent possible modules identified in the ontologies that cope, partially or totally, with the HeC levels, while arrows represent connections. The connections between modules of the same ontology are easy to establish because they are defined during the modularization; whereas the connections between modules of different ontologies involves a complex mapping process.

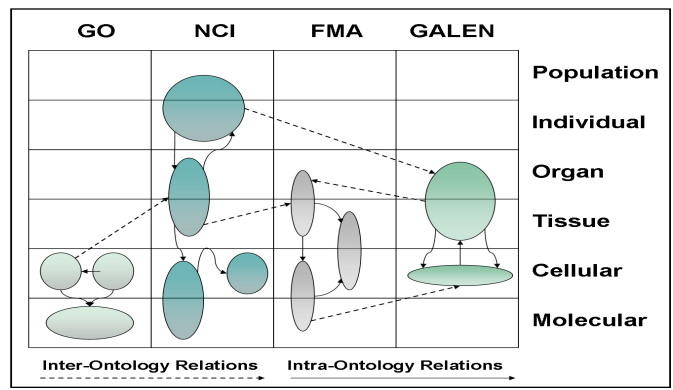

**Fig. 2.** Vertical levels and modules in ontology representations

Classical mapping discovery processes only try to find similarities between two ontologies, by determining which concepts and properties represent the same reality (so-called syntactic and lexical matching) [11]. However modern approaches, such as C-OWL [12] or E-Connections [13], try to find more complex relations (or bridges) between concepts (e.g.: causes-disease, located-at, encodes, involves, etc.). These works refer to these complex relations as *bridge rules* or *E-Connection*, respectively, and infer them by means of external sources (i.e. document repositories such as PubMed, Meta-thesaurus, etc.). In the approach presented in this paper the definition of bridges is mainly based on an earlier work [14] in which a technique for automatically generating ontology instances from texts was applied. The extracted instances not only populate the ontology but also should yield some additional information, potentially useful for completing and refining the ontology definition, or for adding new semantic relations between concepts (semantic bridges). To extract such bridges, for the biomedical domain, public biomedical databases like UNIPROT, OMIM, EMBL, etc. are mined.

## 7 Conclusions

In this paper we have presented a novel methodology for the integration of biomedical knowledge. It specifically addresses vertical integration over diverse granularity levels and describes several techniques to enforce the methodology. Text mining facilities are used to automatically populate ontology instances, providing complementary information for completing the ontology definition and discovered bridges. Semantic bridges are the key to integration and discovery of new knowledge. We believe these powerful concepts will drive us towards the construction of an integrated view of child's health in the European Health-e-Child project.

**Acknowledgements**

The authors wish to acknowledge the support provided by all the members of the Health-e-Child (IST 2004-027749) consortium in the preparation of this paper. This work has been also partially funded by the UJI-Bancaixa project P1B2004-30 and the Spanish Ministry of Education and Science project TIN2005-09098-C05-04.

## References

1. J. Freund, et al. "Health-e-Child: An Integrated Biomedical Platform for Grid-Based Pediatrics", Studies in Health Technology & Informatics # **120,** pp 259-270 IOS Press 2006
2. C. Goble et al. "Transparent Access to Multiple Bioinformatics Information Sources (TAMBIS)". IBM System Journal, **40(2)**, pp.532-551, 2001.
3. J. Kohler, S. Phillipi & M. Lange., "SEMEDA: ontology based semantic integration of biological databases". Bioinformatics **19(18)**, 2003.
4. The Unified Medical Language System (UMLS) fact sheet available at the National Institutes of Health Library of Medicine: http://www.nlm.nih.gov/pubs/factsheets/umls.html
5. D Perez-Roy et al.,"ONTOFUSION: Ontology-based integration of genomic and clinical databases". Computers in Biology & Medicine **36(7-8)**, 2006.
6. R. Stevens, A. Robinson, and C.A. Goble., "myGrid: Personalised Bioinformatics on the Information Grid" Bioinformatics Vol. **19**, Suppl. 1, i302-i304, 2003.
7. I. Oliveira et al., "On the Requirements of Biomedical Information Tools for Health Applications: The INFOGENMED Case Study". BioENG 2003. Lisbon, Portugal.
8. I. Sanz, M. Mesiti, G. Guerrini, R. Berlanga. "ArHeX: An Approximate Retrieval System for Highly Heterogeneous XML Document Collections". Demo at EDBT 2006.
9. I. Sanz, M. Mesiti, G. Guerrini, R. Berlanga. "Highly Heterogeneous XML Collections: How to retrieve precise results?" In Proc. of FQAS. 2006.
10. E. Jiménez, R. Berlanga, I. Sanz, M. J. Aramburu, R. Danger: "OntoPathView: A Simple View Definition Language for the Collaborative Development of Ontologies". Artificial Intelligence Research and Development, pp 429-436. IOS Press, 2005.
11. N. F. Noy. "Semantic Integration: A Survey of Ontology Based Approaches" ACM SIGMOD Record, Special Issue on Semantic integration, 2004.
12. Bouquet, F. Giunchiglia, F. Van Harmelen, L. Serafini, H. Stuckenschmidt. "Contextualizing ontologies", Journal of Web Semantics **1(4)**, pp. 325-343, 2004
13. Cuenca-Grau, B. Parsia, E. Sirin. "Combining OWL Ontologies using E-Connections". Journal of Web Semantics **4 (1)**, pp. 40-59, 2005.
14. R. Danger, R. Berlanga, J. Ruíz-Shulcloper "CRISOL: An approach for automatically populating a Semantic Web from Unstructured Text Collections". In Proc. of DEXA 2004.